# Fundamental Constants at High Energy[1, 2]

**Harald Fritzsch**

Theory Division, CERN, CH–1211 Geneva 23, Switzerland,
and
Ludwig–Maximilians–University Munich, Sektion Physik
Theresienstraße 37, 80333 München

**Abstract:** The progress of Particle Physics is closely linked to the progress in the understanding of the fundamental constants, like the finestructure constant, the mass of the electron or nucleon, or the electroweak mixing angle. The relation between the 18 fundamental constants of the Standard Model and the elementary units used in other fields like quantum optics or solid state physics is far from trivial and will be discussed. Relations between the various constants might exist, providing signals for the physics beyond the Standard Model. Recent observations in astrophysics indicate a slight time variation of the finestructure constant. If true, it has profound implications for many particle and nuclear physics phenomena. In particular the nuclear mass scale should change in time, a phenomenon which could be observed in the laboratory using advanced methods of quantum optics.

The Standard Model of particle physics[1] is a superposition of QCD for the strong interactions and the electroweak gauge theory for the electromagnetic and weak interactions. It gives a nearly complete description of all observed phenomena in atomic, nuclear and particle physics. It is only nearly complete, since certain phenomena, among them the increasing evidence for neutrino oscillations and the dominance of matter as compared to antimatter in the universe, cannot be described within the framework of the Standard Model. The major drawback exhibited by the Standard Model is the fact that a large number of constants, in particular many mass parameters, have to be adjusted according to the experimental measurements and cannot be predicted within the theory. Many theoreticians in particle physics believe for this reason that the Standard Model must be regarded only as a first step towards a more complete understanding, and that in the future it will be embedded in a larger and more complete theoretical framework.

In this talk I cannot offer a solution of the problem of the many fundamental constants, however I shall give a critical overview and describe possible directions one might go. Let me first point out that the Standard Model, which is based on a description of the fundamental forces by the theoretical framework of quantum theory, is a description of the local laws of nature. It does not say anything about the possible boundary conditions

[1] Partially supported by VW–Stiftung Hannover (I-77495)
[2] Invited talk given at the Heisenberg symposium of the Alexander von Humboldt-Foundation (Bamberg, September 2001)

which are imposed on these local laws from the outside, in particular from cosmological boundary conditions. It could well be that at least certain fundamental constants are subject to such boundary conditions. If this is the case, it would not help to find a description of these constants by embedding the Standard Model in a more fundamental theory. Some of the constants appearing in the Standard Model could indeed be cosmic accidents, i. e. quantities which were fluctuating wildly at the time of the creation of the universe, but were frozen immediately afterwards. These constants could be called "frozen accidents"[2]. Other constants could eventually be determined by dynamical laws, which go beyond the laws of the Standard Model. Such constants could indeed be calculated in a future theory. It might also be that some of the elementary constants are not constants at all, but are slowly changing in time. Such ideas were pioneered by Dirac[3], who once proposed that the gravitational constant is a function of the cosmic time. This idea, however, faded away during the course of the last century, since no time variation of the gravitational constant has been observed. Nevertheless one should keep in mind that not only the gravitational constant, but also other constants appearing in the Standard Model might turn out to be slowly varying functions of time.

The constant which plays the most significant role in atomic physics is the finestructure constant $\alpha$ introduced by Arnold Sommerfeld in 1917. Sommerfeld noted that the finestructure of the atomic levels was determined by a dimensionless number whose value today is given by:

$$\alpha^{-1} = 137.03599976(50)\,. \tag{1}$$

Numerically it turned out that the inverse of $\alpha$ is quite close to be an integer number. Sommerfeld himself did not indulge in philosophical speculations about the nature of $\alpha$. Such speculations were started in the thirties by Arthur Eddington, who speculated about an intrinsic relation between the inverse of $\alpha$ and the total number of different charged objects[4]. He introduced a specific counting of these objects, including their spin and came up with the astonishing number 136, which in Eddington's view was sufficiently close to the observed number of 137.

Shortly after Eddington Werner Heisenberg came up with a proposal to describe $\alpha$ by an algebraic formula, which works up to an accuracy of $10^{-4}$:

$$\alpha = 2^{-4}3^{-3}\pi\,, \tag{2}$$

Wyler found in 1971 an algebraic formula based on group–theory agruments which works up to the level of $10^{-6}$[4]:

$$\alpha = \frac{9}{8\pi^4}\left(\frac{\pi^5}{2^4 5!}\right)^{\frac{1}{4}}. \tag{3}$$

Today we must interpret such attemps as useless. In particular in the underlying theory of Quantum Electrodynamics the actual value of the coupling constant changes if one changes the reference point, i. e. changing the energy scale.

The theory of Quantum Electrodynamics is still the most successful theory in science. It brings together Electrodynamics, Quantum Mechanics and Special Relativity. QED is a renormalizable theory and has been tested thus far up to a level of one in 10 Million. Physical quantities like the anomalous magnetic moment of the electron can be calculated in terms of powers of $\alpha$, and $\alpha$ has been determined that way to a high degree of accuracy.

The value of $\alpha$ describes the coupling strength of electrodynamics at distances which are large compared to the Compton wavelength of the electron. At smaller distances $\alpha$ changes slowly. In pure QED, i. e. in the presence of only the photon field and the electron, the effective value of $\alpha$ is given in momentum space by:

$$\alpha_{eff}\left(q^2\right) = \frac{\alpha}{1 - \frac{\alpha}{3\pi}ln\left(\frac{-q^2}{Am_e^2}\right)} \tag{4}$$

$$A = exp\left(5/3\right), -q^2 > 0. \tag{5}$$

The infinitesimal change of $\alpha$ is dictated by the renormalization group equation:

$$\frac{d}{dln(q/M)}e\left(q;e_r\right) = \beta(e), \quad e\left(M;e_r\right) = e_r \tag{6}$$

($M$: renormalization point) .

At high energy not only virtual electron–positron pairs contribute, but also myon pairs, $\tau$ pairs, quark–antiquark–pairs etc. At smaller distances $\alpha$ is becoming larger. This effect can also be seen directly in the experiments. At LEP the effective value of $\alpha$ given at an energy scale of 91 GeV (mass of $Z$–boson) is:

$$\alpha\left(M_z\right) \cong (127.5)^{-1} . \tag{7}$$

The renormalization group requires that the strength of the electromagnetic coupling increases at increasing energy and eventually reaches a point, where perturbation theory breaks down. In pure QED, i. e. in the theory of a photon field, interacting only with one charged fermion, the electron, the critical energy (Landau singularity) is extremely high and far above the energy scale given by gravity. Of course, this theory is not realistic, since in the real world there are 3 charged leptons and 6 charged quarks, and therefore the increase of the coupling constant happens at a much higher rate. Already at energies, which were reached by the LEP–Accelerator, of the order of 200 GeV, the associated value of the finestructure constant is more than 10% higher than at low energy. In any case this signifies that one should not attach a specific fundamental meaning to the numerical value of the finestructure constant.

The fact that in particle physics all phenomena can be described in terms of a number of fundamental constants is, of course, directly related to the Standard Model. The Standard Model is a superposition of the quark–gluon gauge field theory (QCD) and of the electroweak gauge theory, based on the gauge group $SU(2) \times SU(1)$. The Standard Model is not merely a gauge field theory, like many others. It aims at a complete description of all particle physics phenomena, and it is extremely successful in doing so. Let me remind you that the experimental program, using the LEP Accelerator at CERN, came to an end in the year 2000. The outcome of the research done with the LEP Accelerator constitutes a triumph in particular for the standard electroweak gauge theory. The parameters of the theory, most notably the mass of the $Z$–boson and the coupling parameters have been determined with an impressive accuracy.

In the Standard Model the masses of the weak bosons are generated by the coupling of the boson fields to the thus far hypothetical Higgs–field. The model requires the existence of a Higgs particle, whose mass, if it exists, is one of the basic parameters of the Standard Model. The present limit of the mass of the Higgs particle, given by the CERN experiments, is about 110 GeV.

In the Standard Model the number of basic parameters is 18, including the three gauge coupling constants. Thirteen of these constants are directly related to the fermion masses.

The 18 basic constants of the Standard Model can be listed as follows:

$$
\begin{array}{ccccccccc}
m_e & m_u & m_d; & m_\mu & m_c & m_s; & m_\tau & m_t & m_b \\
 & \theta_u & \theta_d & \theta & \delta & & & & \\
 & & M_w & M_h & & & & & \\
\alpha & & \alpha_s & & \alpha_w & & & &
\end{array}
\tag{8}
$$

The nature of most of the fundamental constants seems to be intrinsically related to the generation of masses. One of the peculiar features of the Standard Model is the fact that two different types of mass generation mechanisms seem to operate. On the one hand the masses of the weak bosons and of the fermions are given by the coupling of these fields to the scalar boson. On the other hand the masses of the nucleons and moreover the masses of all nuclei are predominantly due to a dynamical mass generation. The generation of mass in QCD could be described as "mass from no–mass". In lowest order the behavior of the QCD coupling constant $\alpha_s$ is given by:

$$\alpha_s\left(q^2\right) = \frac{2\pi}{b_0 ln\left(\frac{q}{\Lambda}\right)}, \quad b_0 = 11 - \frac{2}{3} n_f \tag{9}$$

($n_f$: number of flavors, $q = \sqrt{q^2}$), $\Lambda$: scale parameter).

Formally the coupling constant becomes infinite, if the energy scale involved approaches the critical value $\Lambda$. Through "dimensional transmutation" the functional dependence of the coupling constant on the energy leads to the appearance of a mass scale. In the limit in which all the quark masses are set to zero, the masses of the bound states (nucleons etc.) are proportional to $\Lambda$. Using the experimental value $\alpha_s(M_z) = 0.1184 \pm 0.0031$, as given by the LEP–experiments, one obtains $\Lambda = 213 + 38/ - 35$ MeV[5].

In principle the nucleon mass, one of the fundamental parameters of atomic physics, can be calculated in terms of $\Lambda$, if the effects of the quark masses are neglected. Thus far an exact determination of the nucleon mass in terms of $\Lambda$ has not been possible, due to the complexity of the calculations, e. g. within the approach of lattice QCD. However, simpler quantities, for example the pion decay constant, have been calculated with success. The pion decay constant is given by the matrix element of the axial vector current:

$$< 0|A_\mu|\pi> = ip_\mu F_\pi . \tag{10}$$

It has the dimension of mass. The theoretical result is[6]:

$$F_\pi/\Lambda = 0.56 \pm 0.05 \tag{11}$$

while the experiments give:

$$F_\pi/\Lambda = 0.62 \pm 0.10 . \tag{12}$$

The good agreement between experiment and theory indicates that QCD is able to describe not only perturbative features of the strong interaction physics, but also indicates that in the future one might be able to calculate more complicated quantities like the nucleon mass with a good precision.

One must keep in mind that the quark masses are non–zero and will influence the numerical value of the nucleon mass. Unfortunately the uncertainties imposed by our ignorance about the contribution of the quark mass terms to the nucleon mass is high. The matrix element of the non–strange quark mass term, the $\sigma$–term, is only poorly known:

$$< p|m_u \bar{u}u + m_d \bar{d}d|p > \approx 45 MeV \pm 25\% . \tag{13}$$

Also the mass term of the strange quarks plays an important role for the nucleon mass. Typical estimates give:
$$< p|m_s \bar{s}s|p > \sim 40\,MeV\,, \tag{14}$$
with an error which is not less than about 50%. Note that the $u-d$ contribution and the $s$–contribution to the nucleon mass are of similar order.

In general we can say that the nucleon mass is a dual entity. The dominant part of it (about 90%) is due to the dynamical mechanism offered uniquely by QCD, i. e. due to the field energy of the confined quarks and gluons. About 10% of the nucleon mass arises due to the nonvanishing masses of the $u, d$ and $s$–quarks. The strange part of this contribution is about as large as the non–strange part. Moreover there is a small electromagnetic term of about 2% (of order $\alpha \cdot \Lambda$).

In particle physics we are confronted at the beginning of the new millenium with the unsolved problem of the spectrum of the lepton and quark masses. The mass eigenvalues show a remarkable mass hierarchy. As an example I mention the masses of the charge 2/3 quarks: $u:c:t \approx 5:1150:174000$ (the masses are given in MeV).

Most of the quark masses and all of the lepton masses are much smaller than the mass scale of the weak interactions given in the Standard Model by the vacuum expectation value $v$ of the scalar field $v \approx 246$ GeV. Only the mass of the $t$–quark is of the same order of magnitude as the weak interaction mass scale. It is remarkable that the mass of the $t$–quark is within the allowed errors equal to the vacuum expectation value divided by $\sqrt{2}$:
$$v/\sqrt{2} \approx 174\,GeV = m_t\,. \tag{15}$$
Such a mass relation might be a hint towards an interpretation as a Clebsch–Gordan–relation, related to an internal symmetry. However, no such symmetry has been identified thus far, and the question remains whether the relation above is an accident or not. Another interesting feature of the quark mass spectrum is the fact that for each charge channel the mass ratios seem to be universal:
$$\begin{aligned} m_d : m_s &= m_s : m_b \\ m_u : m_c &= m_c : m_t\,. \end{aligned} \tag{16}$$
Again a deeper understanding of this scaling feature is missing.

In the Standard Model the transitions between the various families of quarks (and possibly also of the leptons) arise because the states entering the weak interactions are not identical to the mass eigenstates. The transition strengths are in general given by complex amplitudes, however, it is well–known that the multitude of the flavor transitions is given by 3 mixing angles, which I like to denote by $\theta$, $\theta_u$ and $\theta_d$, and a complex phase parameter $\delta$. All transition strengths can be expressed in terms of these four parameters. For example, the Cabbibo transition between the up quark and the strange quark, often denoted as $V_{12}$, is given in the complex plane by $\theta_u$, $\theta_d$ and the phase $\delta$, which is the relative phase between the two angles
$$V_{12} \cong \Theta_u - \Theta_d e^{-i\delta}\,. \tag{17}$$
In the complex plane $V_{12}$, $\theta_u$ and $\theta_d$ form a triangle, which is congruent to the so called "unitarity triangle". Since the absolute value of $V_{12}$ is given with very high precision, a good determination of the angles $\theta_u$ and $\theta_d$ would allow us to determine the shape of the triangle. Thus far only one of the angles of the triangle, denoted usually by $\beta$, has been determined by the experiments, since it is related to the observed strength of the $CP$ violation in the decay of $B$–mesons. However, the allowed ranges are still large: $sin 2\beta \approx 0.45\ldots 1$.

One can show that the angle $\Theta_u$, which describes essentially the mixing between the $u$– and the $c$–quarks is essentially 0 in the limit $m_u \to 0$. Likewise $\Theta_d$ is essentially 0 for $m_d \to 0$. In simple models for the mass generation based on symmetries beyond the Standard Model one finds simple relations between the mass eigenvalues and the mixing angles[7].

$$tan\,\Theta_u \approx \sqrt{\frac{m_u}{m_c}} \qquad tan\,\Theta_d \approx \sqrt{\frac{m_d}{m_s}}. \qquad (18)$$

If these relations hold the unitarity triangle is determined with rather high precision. In particular the angle $\alpha$ which is equal to the phase parameter $\delta$ is essentially $\frac{\pi}{2}$, which would imply the $CP$ violation in nature to be maximal[7]. Relations between the mass eigenvalues and the mixing angles are of high interest since such relations would reduce the number of fundamental parameters of the Standard Model. It is conceivable that all the 3 mixing angles as well as the phase parameters are fixed by such relations, although the exact structure of the relations is still unclear. Further relations, in particular mass relations between the leptons and quarks and relations among the coupling constants can be obtained if the Standard Model is viewed as a low energy limit of a grand unified theory, based on large symmetry groups, e. g. $SO(10)$.

The coupling constants of the gauge groups $SU(3)$ and $SU(2)$ both decrease at high energies, while the coupling constant of the $U(1)$–sector decreases. At very high energies they become of comparable magnitude. If one uses the observed magnitudes of the coupling constants, one finds that they do converge at high energies, however do not meet exactly at one point. The energy scale were they approach each other is about $10^{15}$ GeV. If the gauge groups of the Standard Model are indeed subgroups of a bigger symmetry group and if the symmetry breaking of the grand unified theory happens at one specified energy, one would expect that the three coupling constants meet at one particular point on the energy scale. One but not the only possibility to reach a convergence of the coupling constants is to introduce supersymmetry. In supersymmetry for each fermion of the Standard Model a corresponding boson is introduced, and each boson of the Standard Model is accompanied by a corresponding fermion. Since the partners of the fermions and bosons have not been observed, their masses must be sufficiently high, typically above about 200 GeV.

The supersymmetric partners of the fermions and bosons do contribute to the renormalization of the coupling constants. If one choses a symmetry breaking for the supersymmetry at an energy scale of the order of about one TeV, one finds that the two coupling constants converge at an energy scale of $1.5\cdot 10^{16}$ GeV. In such a theory the three different coupling constants for the strong, electromagnetic and weak interactions are fixed just by one coupling constant, the unified gauge coupling constant at high energies.

In grand unified theories one typically finds also a parellelism between the quarks of charge $-\frac{1}{3}$ and the charged leptons, implying that at the grand unified energy scale the mass of the charged lepton and of the corresponding quark should be equal, e. g.: $m_b = m_\tau$. Indeed, such a relation works quite well for the b–$\tau$–system. The observed fact that the $b$–quark mass is about a factor of 3 larger than the $\tau$–lepton mass comes from the renormalization effect, mostly due to the QCD interaction. Similar relations between the $\mu$–mass and the $s$–quark mass or between the electron mass and the $d$–mass do not seem to hold. The relations between these masses must be more complicated than the one given above. Nevertheless it is conceivable that such mass relations exist.

Taking into account the relations between the fundamental parameters of the Standard Model discussed above, one may ask how many independent parameters might finally remain. The most optimistic answer is 7: one coupling constant for the unified interaction, the 3 masses of the charged leptons and the 3 masses of the charge $\frac{2}{3}$ quarks. Note that the $t$–mass is supposed to describe also the energy scale of the weak interaction, fixing at the same time the $W$– $Z$–masses and the mass of the scalar boson. The strength of the unified coupling constant can be related to the scale parameter $\Lambda$ of QCD.

Gravity does not have a place in the Standard Model. The gravitational interaction is characterized by a critical energy scale, the Planck–mass: $\Lambda_p = 1.221047 \times 10^{19}$ GeV. The interplay between the gravitational interaction and the Standard Model gauge interactions can be described by dimensionless ratios like:

$$\Lambda/\Lambda_p = 0.17 \times 10^{-19}. \tag{19}$$

Such ratios are not fixed by the considerations made above. They are candidates for a time variation on a cosmological time scale. Furthermore the unified coupling constant might also depend on the time, implying that the finestructure constant $\alpha$ becomes a function of time.

Recently one has found indications that the finestructure constant was perhaps smaller in the past. Studying the finestructure of various lines in distant gas clouds, one found[8]: $\delta\alpha/\alpha = (-0.72 \pm 0.18) \cdot 10^{-5}$. It remains to be seen whether this effect holds up in future observations. If the finestructure constant $\alpha$ undergoes a cosmological shift, one should expect similar shifts also to affect the strong interaction coupling constant, in other words the $\Lambda$–scale, which in turn would affect the magnitude of the proton mass. Furthermore the neutron–proton–mass difference, which has an electromagnetic contribution, would change. This would be important for the nucleosynthesis of the light elements. Thus far a systematic study of all effects of a time variation of $\alpha$ has not been carried out.

If one takes the idea of a Grand Unification of the gauge forces seriously, a time shift of $\alpha$ would make sense only if the unified coupling constant undergoes a time shift as well. But this would imply, as recently pointed out[9] that the QCD scale $\Lambda$ would also change in time. As a result the nucleon mass and all nuclear mass scales would be time–dependent. Grand unification implies that the relative change of the nucleon mass is about two orders of magnitude larger than the relative change of $\alpha$. Using advanced methods of quantum optics, a time variation of $\alpha$ and of the nuclear mass scale could be observed by monitoring the atomic finestructure and molecular rotational or vibrational frequencies[9].

In general we can expect that the problem of the fundamental constants of the Standard Model will remain in the focus of research in particle physics at least for the ten years. The success of any new direction in theoretical research should be measured in terms of its power to make predictions about the fundamental constants or about relations among them.

# References

[1] See e. g.: C. Quigg, Gauge Theories of the Strong, Weak and Electromagnetic Interactions (Advanced Book Classics) (Reprint Series); M. Peskin and U. Schroeder, Elementary Particle Physics: Concepts and Phenomena (Texts and Monographs in Physics); O. Nachtmann, A. Lahee and W. Wetzel, Elementary Particle Physics, Berlin Heidelberg (1990)

[2] M. Gell–Mann, private communication

[3] P. Dirac, Nature **192** (1987) 235

[4] For a discussion see: D. Gross, Phys. Today, Vol. 42, Nr. 12 (1989)

[5] S. Bethke, J. Phys. **G26** R27 (2000)

[6] M. Luescher, Phys. Bl. **56** (2000) 65

[7] H. Fritzsch and Z. Xing, Prog. Part. and Nucl. Phys. **45** (2000) 1–81 [hep–ph/9912358]

[8] J.K. Webb et al., Phys. Rev. Lett. **87**, 091301 (2000)

[9] X. Calmet und H. Fritzsch, CERN–TH / 2001–33 [hep–ph 0112110]